\documentclass[aps,pra,preprint,superscriptaddress,groupedaddress]{revtex4-1}
\pdfoutput=1
\usepackage{graphicx}  
\usepackage{dcolumn}   
\usepackage{bm}        
\usepackage{amssymb}   
\usepackage{physics}
\usepackage{tikz}
\usepackage{gensymb}
\usetikzlibrary{shapes,snakes}
\usetikzlibrary{shapes.geometric}
\usepackage{chngcntr}
\usepackage{hyperref}
\usepackage{xcolor}
\usepackage{float}
\usepackage{cleveref}
\usepackage{subcaption}
\usepackage[utf8]{inputenc}
\usepackage[english]{babel}
 \usepackage{amsthm}

\theoremstyle{definition}

\newcommand{\newc}{\newcommand}
\newc{\beq}{\begin{equation}}
\newc{\eeq}{\end{equation}}
\newc{\kt}{\rangle}
\newc{\br}{\langle}
\newc{\beqa}{\begin{eqnarray}}
\newc{\eeqa}{\end{eqnarray}}
\newc{\pr}{\prime}
\newc{\longra}{\longrightarrow}
\newc{\ot}{\otimes}
\newc{\rarrow}{\rightarrow}
\newc{\h}{\hat}
\newc{\bom}{\boldmath}
\newc{\btd}{\bigtriangledown}
\newc{\al}{\alpha_k}
\newc{\be}{\beta_k}
\newc{\ld}{\lambda}
\newc{\sg}{\sigma}
\newc{\p}{\psi}
\newc{\eps}{\epsilon}
\newc{\om}{\omega}
\newc{\mb}{\mbox}
\newc{\tm}{\times}
\newc{\ra}{\rightarrow}
\newc{\non}{\nonumber}
\newc{\ul}{\underline}
\newc{\hs}{\hspace}
\newc{\longla}{\longleftarrow}
\newc{\ts}{\textstyle}
\newc{\f}{\frac}
\newc{\df}{\dfrac}
\newc{\ovl}{\overline}
\newc{\bc}{\begin{center}}
\newc{\ec}{\end{center}}
\newc{\dg}{\dagger}
\newc{\T}{\mathcal{U}}
\newc{\Tp}{\mathcal{V}}
\newc{\J}{\mathsf{J}}
\newc{\sfL}{\mathsf{L}}
\newc{\C}{\mathsf{C}}
\newc{\B}{\mathsf{M}}
\newc{\V}{\mathsf{V}}
\newc{\red}{\textcolor{red}}

\begin{document}

\title{Quantum chaos and macroscopic realism as no-signaling in time}
\author{Manish Ramchander}
\email[]{manishd@imsc.res.in}
\affiliation{Institute of Mathematical Sciences, Taramani, Chennai 600113, India }
\author{Arul Lakshminarayan}
\email[]{arul@physics.iitm.ac.in}
\affiliation{Department of Physics, Indian Institute of Technology Madras, Chennai 600036, India}

	\begin{abstract}
	Macroscopic realism is a set of assumptions about how we experience the world at a classical level. While the Leggett-Garg inequalities are temporal correlations that are violated by quantum systems not obeying such macrorealism, the no-signaling in time condition is also a necessary condition. This compares   measurement outcomes with and without prior measurements. As dynamics and correlations play a central role in these measures, this paper explores the effects of regular versus chaotic dynamics on the violations of macroscopic realism. We observe a close connection between a 3 point out-of-time-order correlator and the conditional probabilities of measurement, and we find unmistakable imprints of chaos on the violations of macrorealism. We provide qualitative semiclassical reasoning for the numerical results involving a kicked top, and for two important initial states that behave very differently.

	\end{abstract}

\maketitle

	\section{Introduction}

	This paper aims to study the impact that chaos has on  macrorealism, which is a set of assumptions proposed to test the limits of quantum mechanics. They codify the intuition of how macroscopic objects behave \cite{Leggett1985,Leggett_2002} and are in direct contradiction to postulates of quantum mechanics when applied to macroscopic systems. These assumptions allow
	derivation of the Leggett-Garg (LG) inequalities and the no-signaling in time (NSIT) condition as found by Kofler and Brukner \cite{KoflerBrukner2013}, which are necessary conditions for macrorealism to hold.
While the Leggett-Garg inequalities are based on the results of sequential measurements in time, the NSIT conditions compare probability distribution of measurement outcomes with or without prior measurements. They also follow from the assumptions of macrorealism and can be violated even when the LG inequality is not.
	
	 Here we employ the NSIT condition to analyze whether quantized chaotic systems could be especially violative against the tests. As macrorealism is a test of correlations in time, it seems particularly fruitful to ask how the violations are sensitive to very different time evolutions, regular versus chaotic. 	 The subject of macrorealism has
	been extensively studied lately in various directions; see \cite{Emary_2013} for a review, as well as \cite{Clemente2015, ClementeKofler2016,Halliwell2017}. Quantum chaos is seeing a surge of activity, with applications from many-body systems to black-hole physics. Newer techniques such as out-of-time-ordered correlators and operator scrambling to the eigenstate thermalization hypothesis have been developed (see \cite{Rigol16} for a review).
	At a foundational level it could play an interesting role in explaining how one classical world emerges from all the unravelings of the decoherent histories of a multiverse \cite{Strasberg_2024}. Chaos, in the sense of a long time averaged positive Lyapunov exponent is seen as a classical property, and the differences between classical and quantum chaos are usually stark, and it is therefore of interest to explore how the assumptions of macrorealism fare in its presence.

	Quoting Leggett from \cite{Leggett_2002}, macrorealism assumes that (1) ``a macroscopic system with two or more macroscopically distinct states available to it will at 
	all times be in one or the other of these states." (2) ``It is possible in principle to determine which of these states the system is in without any effect on the 
	state itself or on the subsequent system dynamics." (3) ``The properties of ensembles are determined exclusively by initial conditions (and in particular not by 
	final conditions)". Kofler and Brukner have show that the NSIT condition, as discussed in detail below, follows from these assumptions \cite{KoflerBrukner2013}. The idea of macroscopic distinctness is not uniquely defined and
	is reviewed in \cite{Florian}. 
	In this article we consider the disturbance due to projective measurements made by Alice on those made by Bob; the two canonical observers. We  work with 
	the well-studied kicked top \cite{Haake1987, Haake, Peres02, Ghose,Chaudhary, Neill16, ArulTop}, a generic Hamiltonian system with an integrable to chaotic transition. It is described by a large spin $j$ object that is subject to both rotation and torsion in orthogonal directions. This is also a Floquet system in that the torsion is applied in periodic kicks, making the chaos possible as the dynamics is an area-preserving map of a sphere onto itself.

    Previous related work in \cite{Kalaga_LG_2021} involves a study of the LG inequality in a quantum nonlinear Kerr-like oscillator externally pumped by a series of
ultrashort coherent pulses. They have shown the changes that accompany a regular-chaos transition in that system. There are some similarities between that study and this, both study a bosonic Floquet system. However, we study a finite dimensional system that is a textbook example of quantum chaos and has many experimental realizations. Most importantly we study a different version of macrorealism, NSIT, rather than the LG inequality. 

	\subsection{Two measures of No-signaling in time}
	
	We want to study how much Alice's measurement can affect Bob's measurement. The question of how much disturbance is there however, 
	immediately begs another question - when were the measurements made? Therefore, the study must necessarily get entangled with discussions of temporal 
	separation between the measurements. 
	Suppose that Alice and Bob measure observables $ \mathbf{J\cdot \hat{a}} $ and $\mathbf{J\cdot \hat{b} }$ respectively, on a kicked top possessing total angular 
	momentum $j$.  Let $t_0$ be the initial time, and let
	Alice make her measurement at time $t_\alpha$ and Bob make his at $t_\beta$, with $t_0<t_\alpha<t_\beta$. 	
	\begin{figure}[H]
		\centering
		\begin{tikzpicture}[ baseline = 3, scale = 1]
		\filldraw
		(0,0) circle (2pt) node[align = left, above] {$t_0$}--
		(3,0) circle (2pt) node[align =center, above] {$t_\alpha$}--
		(8,0) circle (2pt) node[align = right, above]{$t_\beta$};
		\end{tikzpicture}
		\caption{The measurement timeline. Time increases along the axis towards right.}
	\end{figure}
According to macrorealism,  ``a measurement does not
	change the outcome statistics of a later measurement" and is the 
no-signalling in time statement \cite{KoflerBrukner2013}. Its violation immediately implies that Bob's unconditional 
	(when Alice does not measure) and conditional probability (when Alice does measure) distributions are different.

	Let $P(b,a;t_\beta,t_\alpha)$ be the joint probability distribution for Alice's and Bob's measurements described above. 
	Here $b$ is Bob's outcome and $a$ is Alice's; $t_\beta, t_\alpha$ serve as parameters of the distribution, highlighting respective measurement times. 
	Further, we define $ P_B(b ; t_\beta) $ as the unconditional probability for Bob's measurement of eigenvalue $b$ at $t_\beta$ and $P_C(b;t_\beta, t_\alpha)$ 
	as the conditional probability for the same. It follows from the definition that
	\begin{equation}\label{PC}
	P_C(b; t_\beta, t_\alpha ) = \sum_a P(b,a;t_\beta, t_\alpha),
	\end{equation}	
	and NSIT condtion when cast into an equation, becomes 
	\begin{equation}\label{NSIT}
	\forall \;t_\alpha < t_\beta : P_B(b; t_\beta)  = P_C(b; t_\beta, t_\alpha)\qquad \text{(true under MR).}
	\end{equation}
Alice makes only one measurement before Bob, and when Bob eventually compares the probability distributions he gets with and without the measurement, 
	the results may be different. Since a good comparison should focus on both local and global features, we choose the following two complementary measures.

	First is the standard Hellinger distance \cite{vaart_1998}. If we take $N$ ordered pairs of events $(x_i, y_i)$ with $p_i$ ($q_i$)  being the probability corresponding to 
	$x_i$ ($y_i$), then, the Hellinger distance between the distributions is defined as
	\begin{equation}\label{Hellinger}
	H(p, q) = \frac{1}{\sqrt{2}} \sqrt{\sum_{k=1}^N (\sqrt{p_i} - \sqrt{q_i})^2 }.
	\end{equation}
    This distance is bounded between $0$ and $1$. Note that if and only if $p_i = q_i$ for each $i$, it is zero (a condition necessary and sufficient for violation of NSIT) and becomes maximum whenever $p_i = 1, q_j = 1; i\neq j$.

    We consider another measure: the difference in how many values of $\mathbf{J\cdot \hat{b}}$ could Bob have got in the two cases. This measure can be defined using the participation ratio. Suppose that $p$ is a 
	probability mass function for X which can take values $\{x_1, x_2, \ldots x_N\}$. Then \begin{equation}
		W (p) = \left(\sum_{k=1}^N p(x_k)^2\right)^{-1}
	\end{equation} 
	is a good measure of how much the distribution $p$ is spread, being 
	bounded between 1 (when only one value is possible) and $N$ (when all values are equally likely). We can compute it for $P_B$ and $P_C$ to get the accessible states.
	Then we can define
\begin{equation}\label{PPr}	 \Delta(P_C, P_B) = W(P_C) -W(P_B) 
	\end{equation}

Regarding the temporal separation $t_{\beta}-t_{\alpha}$, in general Bob may choose to vary his measurement time from $t_{\beta}=0$ to $\infty$, according to an arbitrary probability 
	distribution. Then, for a fixed $t_\beta$, Alice may choose to measure at any time before Bob, according to a probability distribution of her choice. 
	Simplest examples would be the delta and the uniform distributions for both of them. We restrict the study to the following scenario:
	\begin{itemize}
	\item Alice measures at a fixed time interval before Bob (the delta distribution), and Bob's time is a uniform random variable in a range $(1, T)$. 
	\end{itemize}	
This set up shall capture several features of interest.

\section{Measurements on a kicked top} 	
		
		Quantifying the disturbance in the spirit of NSIT condition, we shall make the said connection between chaos and macrorealism. 
	The Hamiltonian generating the time evolution is taken to be the kicked top:
	\begin{equation}\label{Hamiltonian} H(t) = J_y \frac{\pi}{2} +\frac{\kappa_0 }{2j} J_z^2 \sum_{n=-\infty}^{\infty} \delta (t-n).
	\end{equation}
	Here $\kappa_0$ is a parameter signifying the kick strength. The kicked top displays chaos in the classical limit $j \rightarrow \infty$ as $\kappa_0$ increases 
	beyond $2$ and becomes almost fully chaotic for $\kappa_0>6$. The corresponding quantum unitary map is the Floquet operator connecting states across a 
	time-period (here chosen as unity) is given by 
	\begin{equation}\label{Map}
	U = \exp(-i\kappa_0 J_z^2/2j) \exp(-iJ_y\pi/2)
	\end{equation}
	which we shall often write as $U = TR$ where $T =e^{-i\kappa_0 J_z^2/2j}$ comes from the twist or torsion about the $z$ axis and $R = e^{-iJ_y \pi/2}$ is rotation about the $y$ axis. 
    
    As noted earlier, there is a major body of work concerning this ``simple" dynamical systems, see the books by Haake \cite{Haake} and Peres \cite{Peres02} including a cold atom experiment \cite{Chaudhary}, and a transmon based one \cite{Neill16}. It has been used in studying the impact of chaos on entanglement \cite{ArulTop,Ghose},
    and for small values of $j$ it can be analytically solved and presents many intriguing features \cite{RuebeckArjendu2017,ArulTop}  such as a proto-exponential growth of OTOC \cite{PG_2021}. This system becomes increasingly macroscopic with increasing $j$, and therefore is 
	relevant for the study of macrorealism tests.

	Quantum mechanics intrinsically violates the assumptions of macrorealism. Therefore it is not a surprise that calculations would lead to violations of NSIT, especially for small $j$.
		To answer whether the degree of discord between the two depends on chaos or not, it
	is imperative to focus on identifying direct effects of chaos and filter out intrinsic 
	violations arising from other known sources, such as incompatible measurements. Restricting to special initial states and
	measurement schemes will enable us to do so.
		
		\subsection{The conditional probability, connection to an OTOC, and the role of coherence}
	Here we consider what Alice's measurement does to the system, and discuss  restrictions on $\mathbf{\hat{a},\hat{b}}$, the axes of measurements. 
	Firstly, note that in calculating \eqref{PC}, Alice's action can be considered as a measurement that reduces the original pure state to a mixed state. To see this, let $\rho_{0}$ be the initial state and $\rho_\alpha$ be the state before Alice's measurement and suppose that the projectors of  
	$\mathbf{J\cdot \hat{a}}$ are 
	$\{A_a\} $. Then \begin{equation}\sum_a A_a \rho_\alpha A_a
	\rho_\alpha' = \sum_a A_a \rho_\alpha A_a=\sum_a A_a U_{\alpha} \rho_0 U_{\alpha}^{\dagger} A_a
	\end{equation} 
	is the claimed mixed state post measurement. Here $U_{\alpha}=\exp(-iHt_{\alpha})$ or $U^{t_{\alpha}}$ in the Floquet case. Subsequently, evolution occurs for $t_\beta-t_\alpha$, after which Bob measures $\mathbf{J\cdot \hat{a}}$ whose
	projectors are $\{B_b\} $, and finds the distribution $P_C$:  
	 \begin{equation}
		P_C(b; t_\beta, t_\alpha) = \Tr[U_{\beta\alpha}\rho_\alpha'U_{\beta\alpha}^{\dagger}B_b] = 
		\sum_a \Tr[U_{\beta\alpha} A_a \rho_\alpha A_a U_{\beta \alpha}^{\dagger} B_b ].
        \label{eq:PC}
	\end{equation}
	We use $U_{\beta\alpha}=\exp(-iH(t_{\beta}-t_{\alpha}))$ for the evolution operator from $t_{\alpha}$ to $t_{\beta}>t_{\alpha}$. 
    In the case of Floquet operator $U$ as we use in this work $U_{\beta\alpha}=U^{t_{\beta}-t_{\alpha}}$, where these times are integers.
    Multiply and divide by $\Tr[\rho_\alpha A_a]$ for each term in the sum, to get \[
	P_C(b; t_\beta, t_\alpha) =	\sum_a \Tr[U_{\beta\alpha} \rho_a U_{\beta \alpha}^{\dagger} B_b] \Tr[ \rho_\alpha A_a ]
	\] 
	where $\rho_a = A_a \rho_\alpha A_a/\Tr[\rho_\alpha A_a]$. This is of course $\sum_a P\left(b, a ; t_\beta, t_\alpha  \right) $, and is the same as \eqref{PC}. The unconditional probability is simply $P_B(b;t_{\beta},t_{\alpha})=\Tr\left( U_{\beta} \rho_0 U_{\beta}^{\dagger} B_b\right)$.

    We now observe that the conditional probability is one the simplest forms of an out-of-time-ordered correlator or OTOC. Let $U_{\alpha}^{\dagger} A U_{\alpha}=A(t_{\alpha})$ be the Heisenberg evolution of the projector over the times $t_{\alpha}$. Using such Heisenberg evolved projection operators, it is simple to then rewrite Eq.~(\ref{eq:PC}) as
    \begin{equation}
        P_C(b; t_\beta, t_\alpha) = \Tr\left[\rho_0 A_a(t_{\alpha}) B_b(t_{\beta})A_a(t_{\alpha})\right]. 
        \label{eq:3OTOC}
    \end{equation}
    This is the simplest form of an OTOC as the times are ordered non-monotonically. It has been referred to as a 3-OTOC and used in \cite{Hamazaki_OTOC_2018,Jalabert_OTOC_2018} where it has been shown it play an important role in the growth of the norm of commutator. This quantity has not been nearly as much studied as the 4-point OTOC, but it is clear that this is related to the conditional probability. In fact OTOC has also been studied as a measure of ``disturbance" and has been pressed into service to elucidate the so-called ``quantum butterfly effect", see \cite{Swingle_2024} for a recent review. 
    
	Next, we note that measurements of non-commuting operators $\vb{J\cdot \hat{a}}, \vb{J\cdot \hat{b}}$ will give violations that typically would get amplified in time, but may have nothing to do with the dynamics itself. To see this possibility, specialize to no evolution. Let the initial state be an eigenstate of $\vb{J \cdot \hat{b}}$. If $\vb{\hat{a} \neq \hat b}$, Bob's measurement will change the state from what Alice's measurement had reduced it to, and thereby create a difference in $P_B$ and $P_C$. Any bonafide distance we use will give non-zero value in their comparison. Even when system evolves, this effect persists because of continuity. Origin of this violation lies in the fact that $[\vb{J\cdot \hat{a}}, \vb{J \cdot \hat{b}}] \sim \vb{\hat{a}}\times \vb{\hat b}$ and not in chaos.  Avoiding such violation would decrease noise, and hence allow a better understanding of the effects that do arise from chaos.
 	Therefore, we restrict to $\mathbf{\hat{a}=\hat{b}}$ case, and henceforth work with density matrices in the $\mathbf{J\cdot \hat{a}}$  eigenbasis $\{|\mathbf{\hat{a}}, m\rangle \}$.

    As there is now a preferred basis for both the measurements, the density matrix is best expressed in this basis. The origin of violations for such cases lies in the removal of off-diagonal terms from the density matrix, because of Alice's
	measurement.   
    To see this, suppose $\rho_\alpha = \sum_{m,n} \varrho^{mn}_\alpha |\mathbf{\hat{a}}, m \rangle\langle \mathbf{\hat{a}}, n| $ is the expansion of the state 
	before Alice's measurement and let $A_k = |\mathbf{\hat{a}},k \rangle\langle \mathbf{\hat{a}}, k|$. Then post measurement, 
	\begin{align}
		\rho_\alpha' = \sum_k A_k \rho_\alpha A_k  &= \sum_{m,n,k} \varrho^{ nm}_\alpha \delta_{km}\delta_{nk}|\mathbf{\hat{a}}, k \rangle\langle \mathbf{\hat{a}},k| = \sum_{k} \varrho^{k k}_\alpha A_k.
	\end{align}
 	Note that the diagonal terms are unchanged. Therefore, deletion of off-diagonals leads to all the observed differences between $P_B$ and $P_C$.  As time evolves, these non-diagonal
	entries effect the probabilities of measurement, because states evolves via conjugation by $U$. The off-diagonal elements measure the quantum coherence of the state, and this is a known resource for quantum information. While measurement destroys this coherence, dynamics restores it, and the NIST is a measure of this competition.
    
    Motivated by this, we consider the $l_1$ norm of coherence  \cite{Baumgratz} \begin{equation}\label{DistPower}
		C_{A}(t_\alpha^-) = \sum_{k\neq m} |\varrho^{km}_\alpha|		
	\end{equation}
	as a measure of disturbing power, because it measures strength of the off-diagonals in the matrix at time $t_\alpha^-$, just before Alice's measurement. Here $A$ in the subscript denotes the basis chosen by her.
	When it is zero, Alice causes no disturbance, and when it is large, intuitively, her measurement becomes more destructive. As $\kappa_0$ 
increases, $C_{A}$ accentuates for initially localized
states because of the non-trivial action of twist operator $T$ (discussed in appendix). This leads to increased violation
due to Alice's measurement, which is seen in the results below. 

	\subsection{Role of time between measurements} 

    As we are dealing with a discrete time dynamics, all times such as $t_{\alpha}$ and $t_{\beta}$ are now integers.
	Alice and Bob decide on $t_\beta-t_\alpha=n$, where Bob's time of 
	measurement $t_\beta$ is a uniform random variable over the set 
	$\{n, n+1,\cdots ,n+T\} $. As a result, any distance computed between $P_B$ and 
	$P_C$ which depends on $t_\beta$ is another 
	random variable.  For a distance function  $d(t_\beta,t_\alpha)$, we define
    $d_n(t_\beta)\equiv d(t_\beta, t_\beta-n)$, whose average is given by  \footnote{For a uniform random variable $t_\beta$, such a time average correctly gives the ensemble average for $d_n$. Same holds for higher moments.}: \begin{equation}\label{key}
	\expval{d_n} = \frac{1}{T+1} \sum_{t_{\alpha}=0}^{T} d_n(n+t_{\alpha}).
	\end{equation} 
	Here $d$ may be $H$ or $\Delta$, with $T\gg 1$.  In the following, we study this average as a function of $\kappa_0$ for two special initial states which are coherent states.
    In the kicked top  the coherent state $\ket{\vb{\hat{y}},j}$ corresponds to a fixed point in the classical 
	map, and $\ket{\vb{\hat{z}}, j}$, is part of a 
	period-4 cycle. 
    
    Being coherent \cite{Glauber, Radcliffe_1971}, they 
	come closest to points in the phase space, which 
	are classical states. Further,
	they both display quantum signatures of chaos \cite{ArulTop} when corresponding classical 
	orbits lose stability with increase in $\kappa_0$. 
	Therefore, they are good examples for studying effects of chaos in $\Delta$ and $H$. We push their further discussion to the appendix for clarity. We set $\mathbf{\hat{a}=\hat{b}=\hat{z}}$ in the following and also $j=15$ as a reasonably large system size. We briefly discuss the generalization for other values of $j$.  

%
	\subsubsection*{Even-odd $n$ differences in $J_z$ measurements}
\begin{figure}[h!]
		\centering
		\includegraphics[scale=.5]{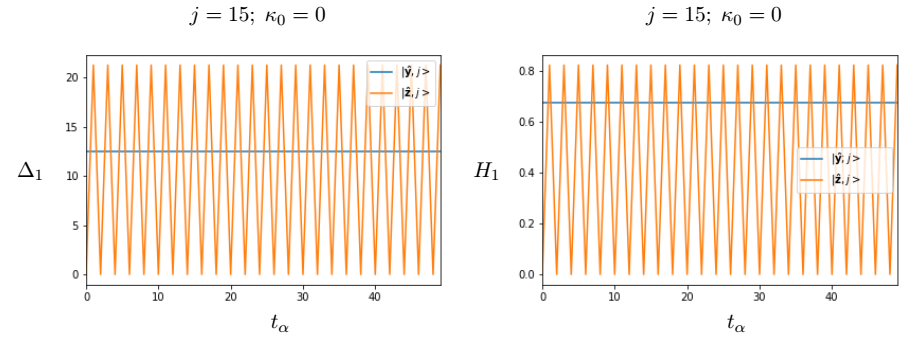}
		\caption{The distances $\Delta_1$ and $H_1$, for $n=1$ in the absence of chaos: $\kappa_0=0$, as a function of Alice's measurement time $t_\alpha$. The constant line is for the initial state $|\vb{\hat{y}}, j\rangle$, while the oscillatory curvem vanishing for even $t_{\alpha}$ is for the initial state $|\vb{\hat{z}},j\rangle $.  Due to continuity,	this effect persists for small non-zero values of $\kappa_0$ as well.   }
		\label{noise}
	\end{figure}

 If Alice and Bob both measure $J_z$, the violations recorded
 in $\Delta$ and $H$ are critically dependent on whether $n$, the time between measurements, is
 even or odd,  especially at small values of $\kappa_0$. As an extreme case, at $\kappa_0=0$, we see that for odd $n$, $\Delta$ and $H$ are non-zero (Fig.~ \ref{noise} shows the case for $\Delta$), whereas for even $n$, both are zero.
	 This happens because of the specific initial states and 
	 measurement axis. When $\kappa_0=0$ the state $|\mathbf{\hat{z}},j\rangle$ evolves to give the 4-cycle as the dynamics is simply $R$, rotation by $\pi/2$ about the $y-$axis: \begin{equation}\label{key}
	\ket{\vb{\hat{z}}, j} \rightarrow \ket{\vb{\hat{x}}, j} \rightarrow \ket{\vb{\hat{z}}, -j} \rightarrow \ket{\vb{\hat{x}}, -j} \qquad (\text{repeat}).
	\end{equation}
	At even $t_\alpha$, state returns to a $J_z$ eigenstate and hence Alice's measurement creates no effect. These cases
	correspond to zero values seen in Fig.~\ref{noise}. At odd $t_\alpha$ however, she produces 
	mixture of $J_z$ eigenstates, which is invariant under $R^2$. If then Bob measures at even $t_\beta$, so that $n$
	is odd, what would have been a $J_z$ eigenstate for him 
	without the measurement, becomes a mixture too. These
	cases correspond to the maxima seen in Fig.~\ref{noise}.
	For even $n$ and odd $t_\alpha$,  using the fact that measurement axis is common and that mixtures produced are 
	symmetric under $R^2$, it follows that $\Delta, H=0$. 
	 For initial state $\ket{\vb{\hat{y}}, j}$, there is no evolution 
	for $\kappa_0=0$. Thus, Alice's measurement can only reduce it to one mixed state which is again invariant under $R^2$, but changes under $R$. Again, as 
	a result, odd $n$ give violation, whereas even $n$ do not. This time however, there are no oscillations as $t_\beta$ varies.

	 Therefore even in plain rotation (when there is no chaos), we 
	see significant violation of NSIT for odd $n$ values. This 
	effect, by continuity, persists even when $\kappa_0$ increases, and for the 
	purposes of establishing a connection between chaos and the signaling, it is unwanted. Therefore, we turn out attention to only even
	$n$ for both the states.
    
	\subsection{Results for initial states $|\mathbf{\hat{z}},j \rangle $ and $\ket{\mathbf{\hat{y}}, j}$ when $\kappa_0 \neq 0$}	
	
	Figure \ref{z,j} shows results for the average of $\Delta_n$ and $H_n$ when the initial state is $|\mathbf{\hat{z}},j \rangle $ for a few values of even $n$.  It is seen that a nearly linear growth is observed in both quantities with a growing slope as $n$ increases.
    The NSIT violation plateaus in both quantities when $\kappa_0$ is between $3$ and $4$, beyond 
	which the variations between different $n$ are lost. Physically, it means
	that if there is strong mixing, on average it doesn't matter how long 
	ago Alice measured on the system. Her measurement causes an equally powerful effect, provided, she measures even periods before Bob.
    The disturbance due to Alice's measurement leads to an enlargement in Bob's outcome possibilities, on an average. As a result we observe that $\langle W(P_C) \rangle > \langle W(P_B)\rangle $, this in fact is seen in all the cases we have studied and needs further exploration. The 4-period cycle corresponding to this state undergoes a change of stability at $\kappa_0=\pi$. We identify
	the saturation-like behavior in $\kappa_0$ as a consequence of this loss of stability.
\begin{figure}[h!]
\centering
\begin{subfigure}{.5\textwidth}
  \centering
\includegraphics[width=.9\textwidth]{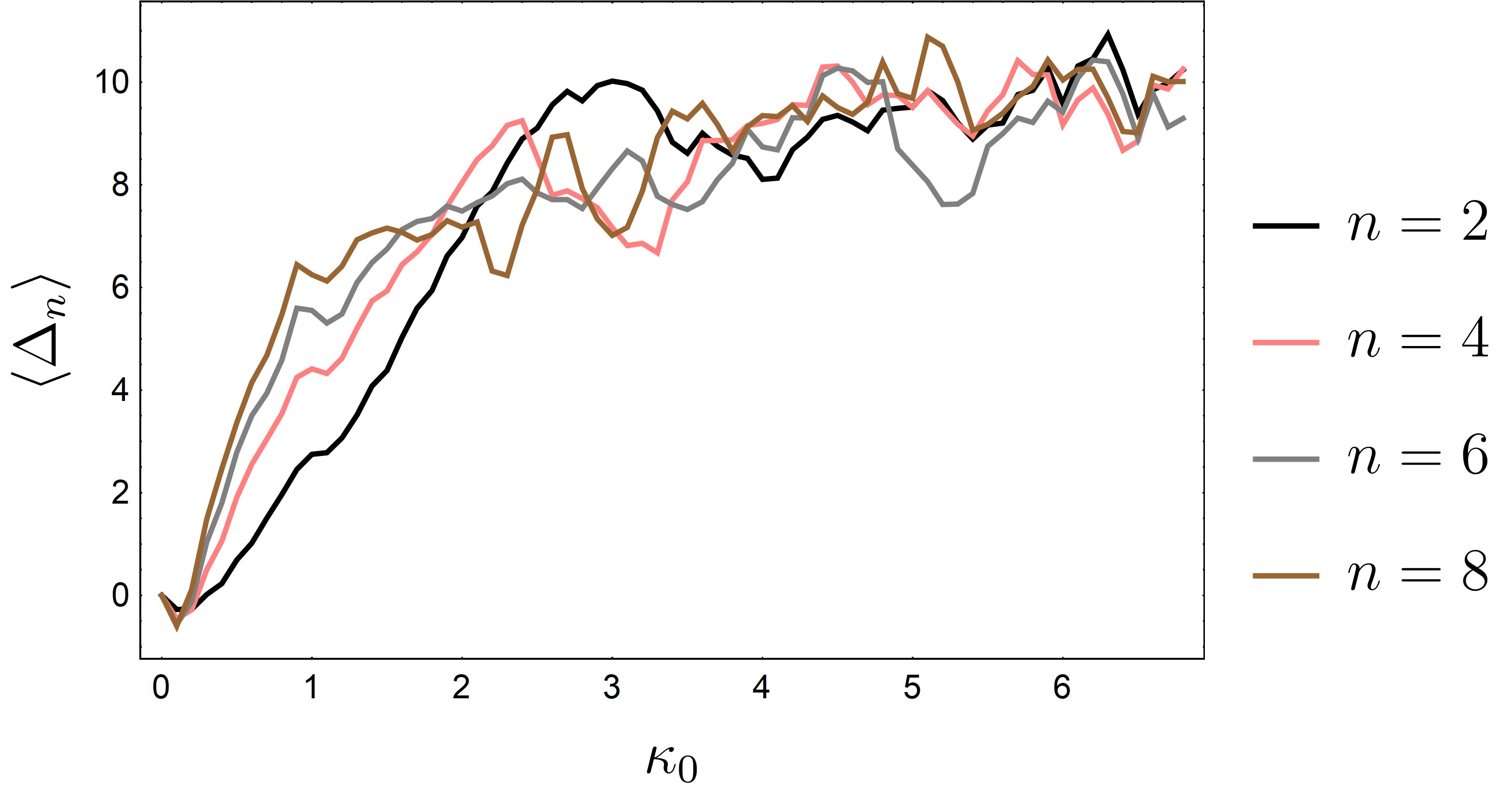}
\end{subfigure}%
\begin{subfigure}{.5\textwidth}
  \centering
\includegraphics[width=.9\textwidth]{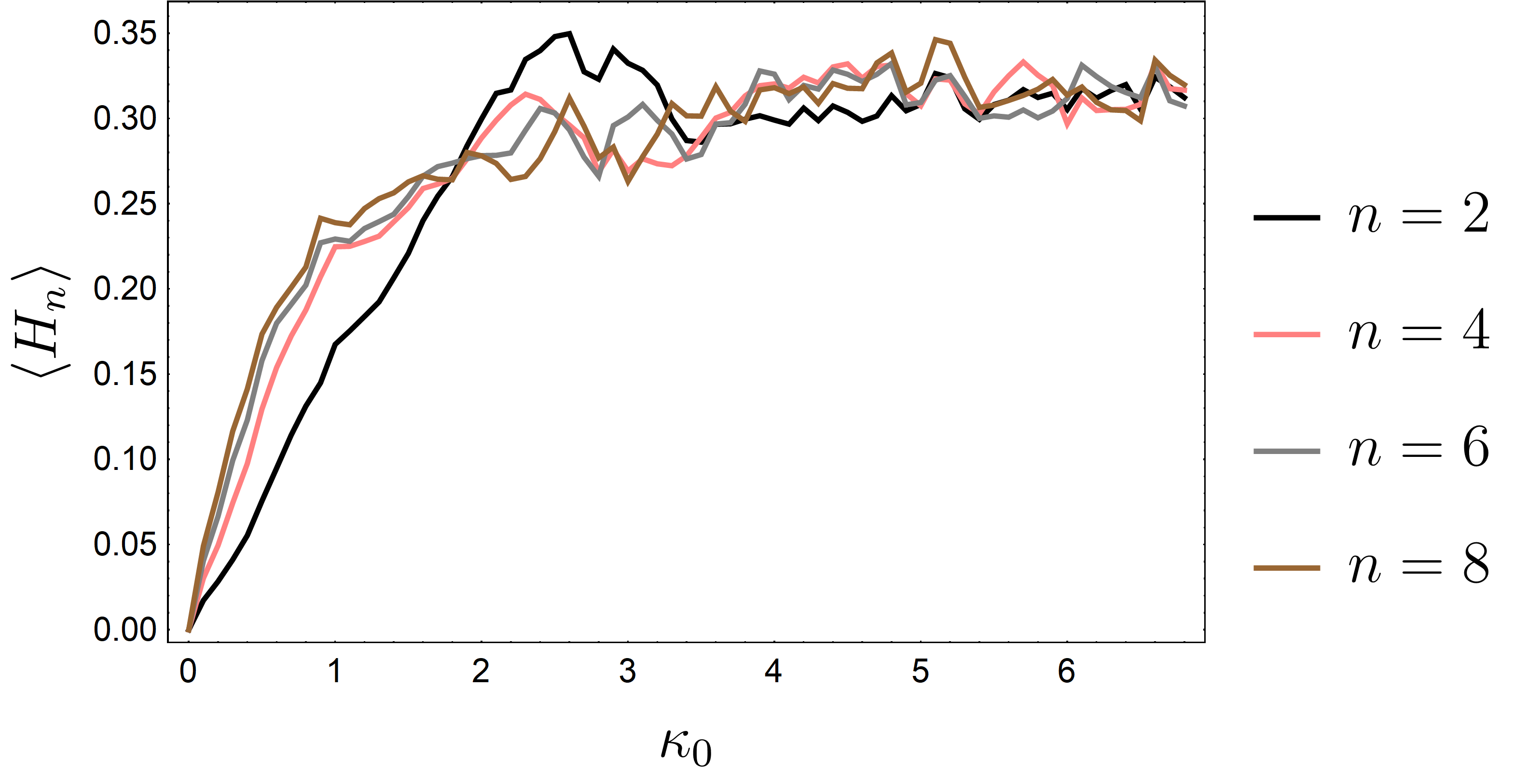}
\end{subfigure}
		\caption{ The distances $\Delta_n$ and $H_n$,
        averaged over $T=50$ initial times $t_\alpha$.  For the initial 
		state  $\op{\vb{\hat z}, j}, j=15$, Alice measures  $J_z$ at $t_{\alpha}$ while Bob measures $J_z$ at $t_{\alpha}+n$. }
	\label{z,j}
	\end{figure}

\begin{figure}[h!]
\centering
\begin{subfigure}{.5\textwidth}
  \centering

\includegraphics[width=.9\textwidth]{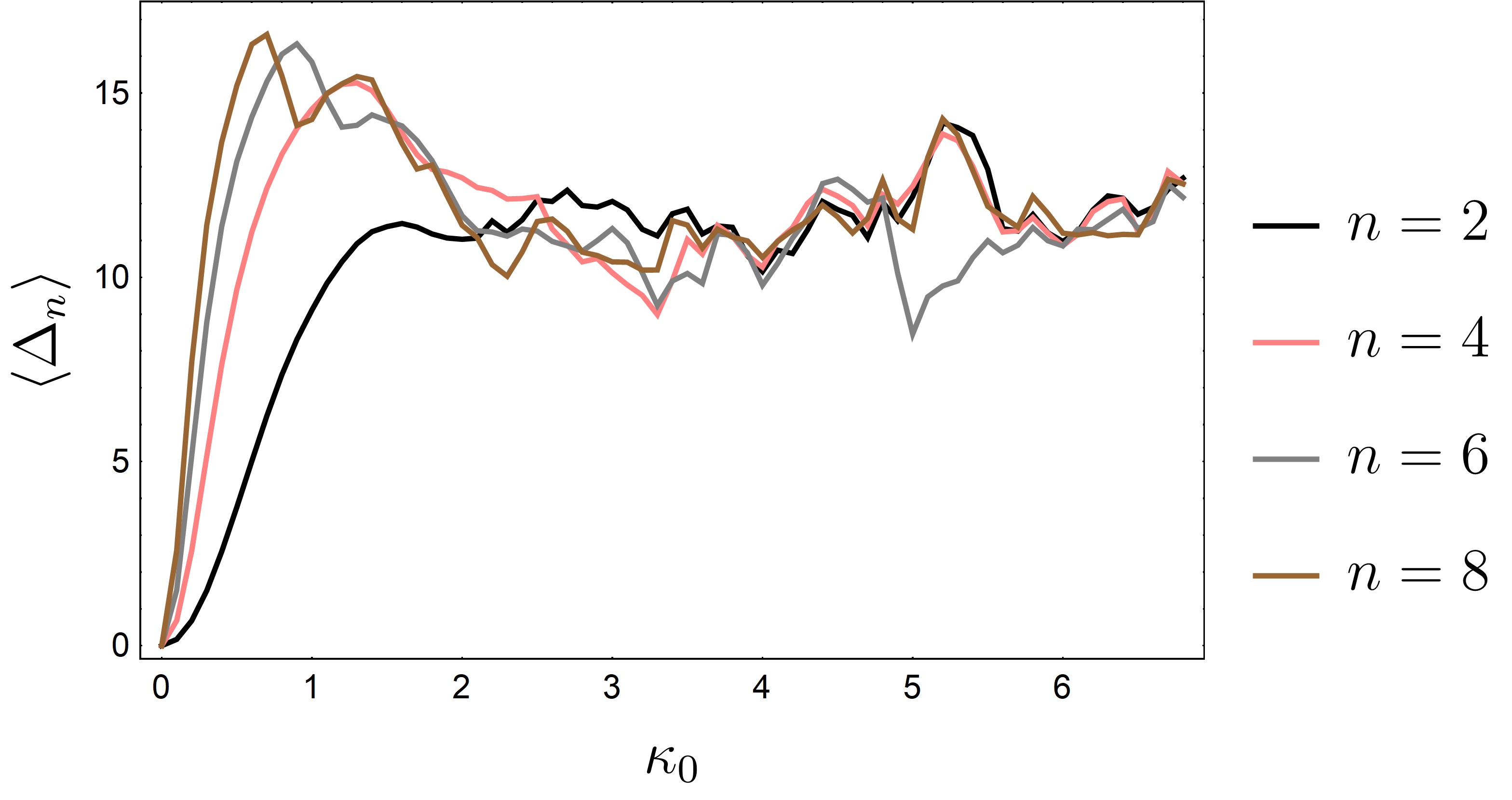}

\end{subfigure}%
\begin{subfigure}{.5\textwidth}
  \centering

\includegraphics[width=.9\textwidth]{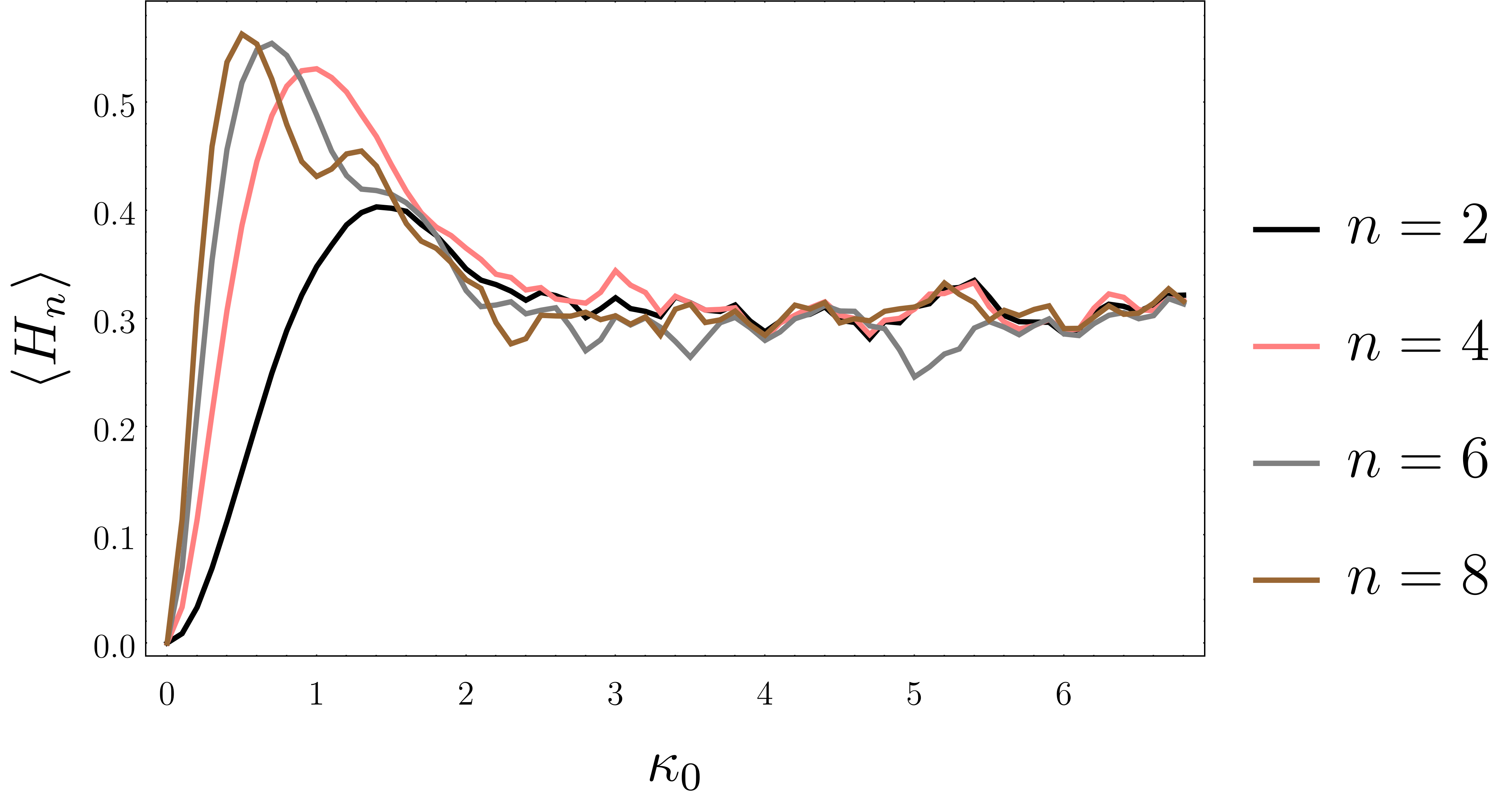}
\end{subfigure}
\caption{Same as the previous figure, except that the initial state is $\rho_0= \op{\vb{\hat y}, j}$. Several features are
	as they were in the case of $| \mathbf{\hat{z}},j\rangle$. Note however, that
	the peaks occur around $\kappa_0=2$ where the fixed point loses its stability. 
	For small $\kappa_0$, the prominent difference for different 
	$n$ values arises because of time evolution for the state localized in the regular region. }
\label{fig:Ystate}
\end{figure}

Figure \ref{fig:Ystate} shows the results when the initial state is $\ket{\vb{\hat{y}}, j}$. A similarity with the $\ket{\vb{\hat{z}}, j}$ case is seen, in as much as there is a growth of the NSIT distance measures as a function of $\kappa_0$ and a saturation that corresponds to the onset of classical chaos. The saturation values are approximately the same. However there are two evident differences. 
(i) The point where the functions saturate is at a smaller value around $\kappa_0=2$. As mentioned the classical point corresponding to the initial state is a fixed point which loses its stability at $\kappa_0=2$, which corresponds well to the parameter at which macrorealism indicators also saturate.
(ii) There is a prominent overshoot of these functions before the saturation. The overshoot gets sharper as the time between the measurements increases.

The second difference is worth discussing further and advance possible causes for it. First, notice that these are averages taken over the first 50 starting times for Alice's measurement from $t_{\alpha}$ from $0$ to $50$, therefore these are rather large violations. 
The maximum violations will occur when Alice's measurement results states that are very different from those produced without her actions. 
That the measurements happen in the $J_z$ basis, implies that the post-measurement states are an ensemble of $J_z$ eigenstates that under the kicked top for small $\kappa_0$ are restricted to the $x-z$ plane, while without Alice's measurement the state would still have been an approximate $J_y$ eigenstate as the initial state is on a stable fixed point. Thus the resultant states fare very different under $J_z$ measurements. This is a qualitative heuristic reasoning that may explain that the larger violations do not seem to happen after the onset of chaos.
    
\begin{figure}[h!]
	\centering

\includegraphics[scale=0.6]{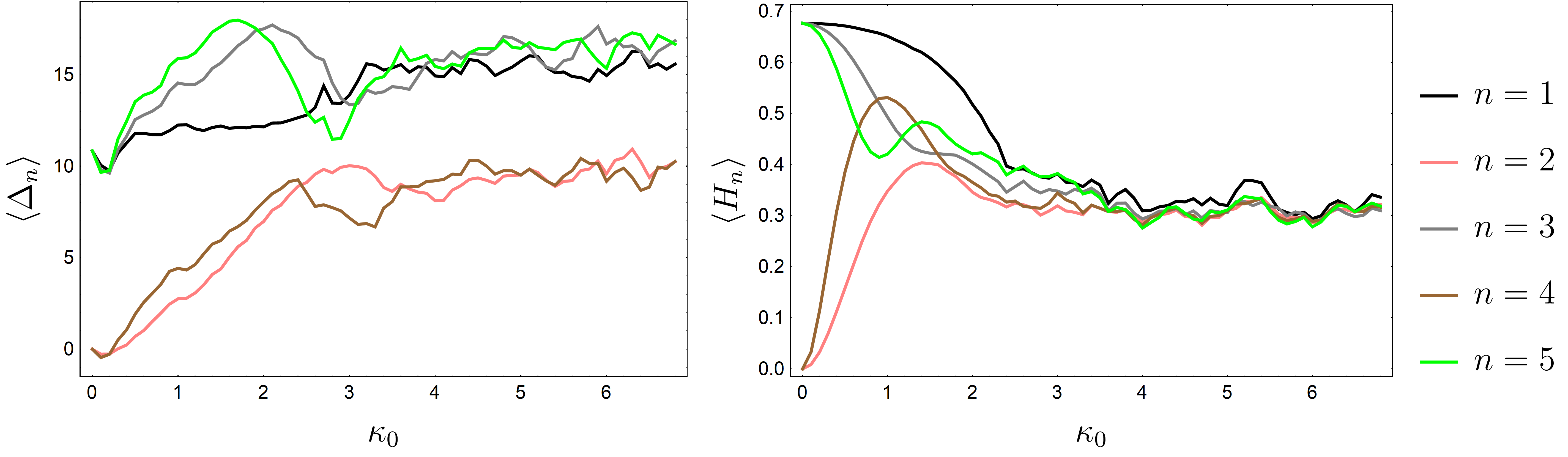}
	\caption{The distance measures are shown including odd and even intervals in the range $1 \leq n \leq 5$.  for the initial state$ |\mathbf{\hat{z}},j\rangle$ (left) the odd/even effect persists well into the fully chaotic regime. For the $|\mathbf{\hat{y}},j\rangle$ case (right), for low $\kappa_0$ values, the proximity to rotation about the $y-$axis produces the difference which  vanishes with the onset of chaos.    }
	\label{oddcases}
\end{figure}

We have not shown so far what happens when $\kappa_0>0$ and the time between Alice's and Bob's measurements is an odd integer. Figure~\ref{oddcases}, shows this as it includes all intervals from 1 to 5. The differences seen 
arise because of the special nature of the states and the axis of 
measurement. For small values of $\kappa_0$ the rotation dominates over the torsion. However, it is interesting and surprising to note that in the case of the initial state being
$|\mathbf{\hat{z}},j\rangle$ the system has not ``forgotten" even when $\kappa_0$ is very large and is deep in fully chaotic regime. This indicates that the type of states that result with measurements and chaos can be quite different from random states. In contrast when the initial state is $|\mathbf{\hat{z}},j\rangle$, odd and even time intervals do not matter when there is global chaos.
Note that in this figure different measures are used in for the two cases, but this is not essential, as they are qualitatively identical.

\begin{figure}[h!]
\centering
\begin{subfigure}{.5\textwidth}
  \centering

 {\includegraphics[width=.9\textwidth]{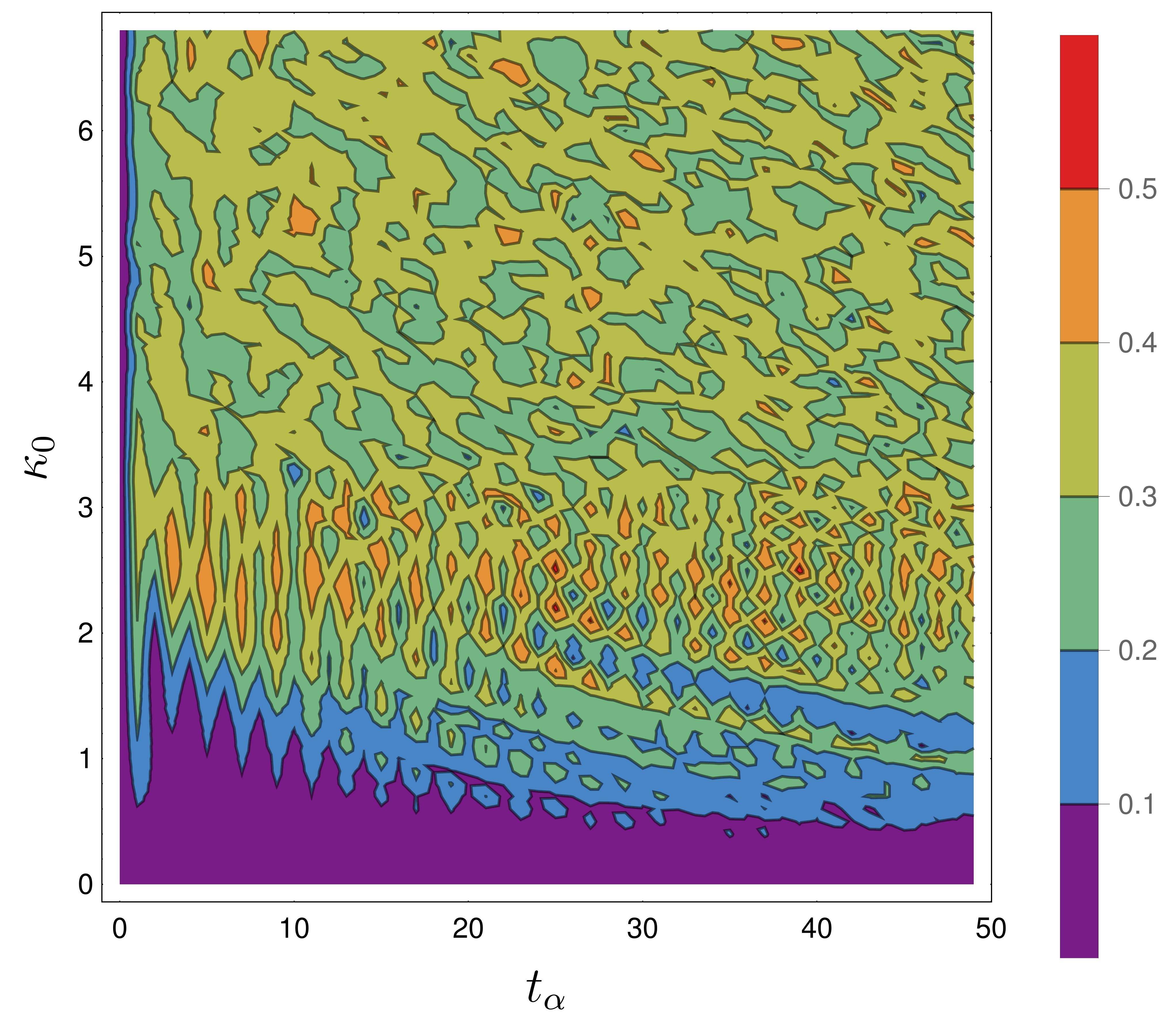}};

  \label{fig:sub1}
\end{subfigure}%
\begin{subfigure}{.5\textwidth}
  \centering

 {\includegraphics[width=.9\textwidth]{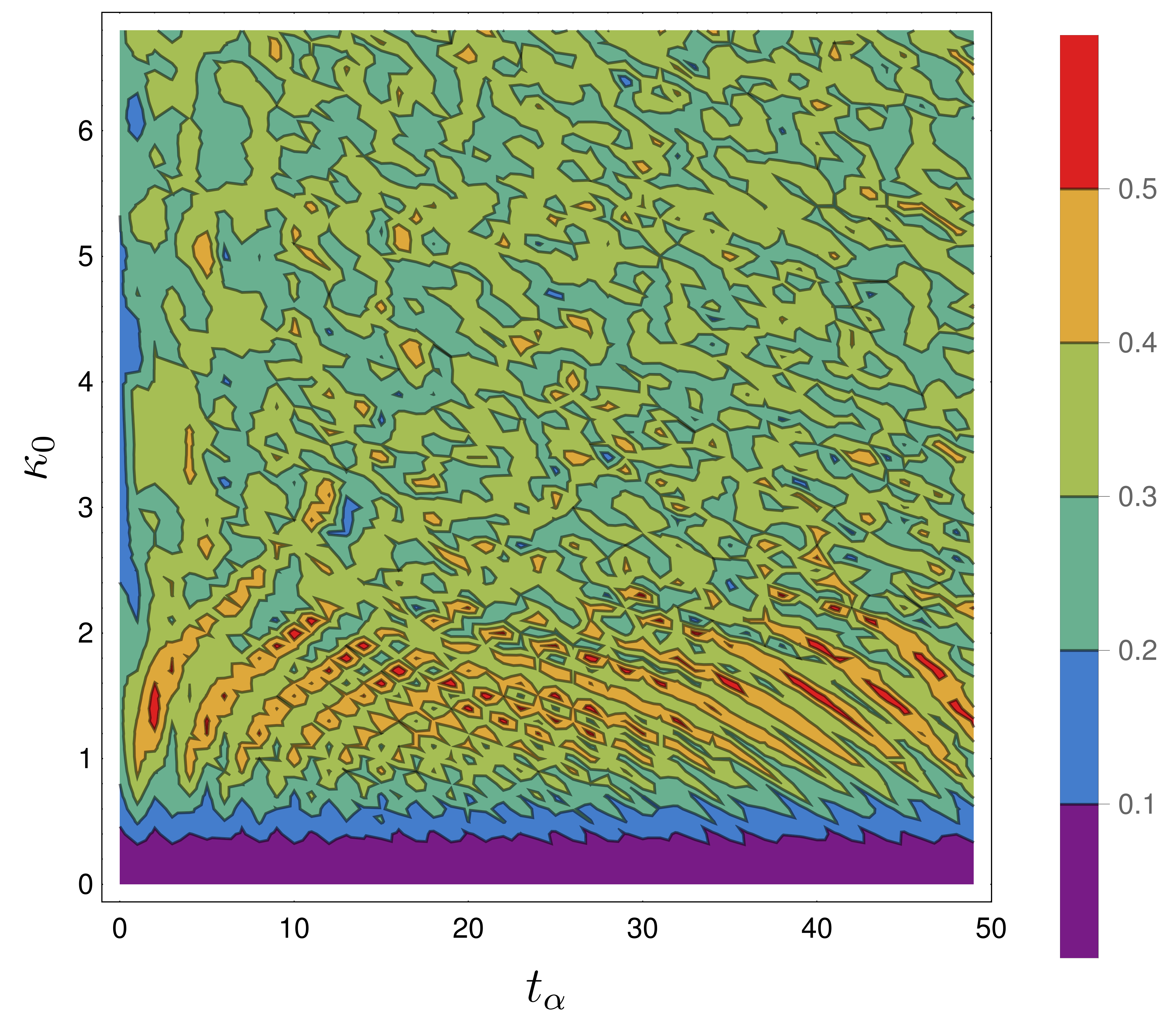}};

  \label{fig:sub2}
\end{subfigure}
\caption{Contour plot of $\Delta_2$, as a function of the time of Alice's measurement, $t_{\alpha}$ and the parameter $\kappa_0$ which controls chaos. Shown are the cases when the initial state is $|\mathbf{\hat{z}},j\rangle$ (left) and $|\mathbf{\hat{y}},j\rangle$ (right).Note the change in behavior of oscillations beyond $\kappa_0 \sim \pi$ on the left, and the change at $\kappa_0 \sim 2$ on the right. The odd-even effects are seen as oscillations that are especially visible in the left plot.}
\label{contour_z}
\end{figure}

Figure \ref{contour_z} shows the contour plots corresponding to
the time averaged plots in figure \ref{z,j;coherence}. Note that in (a)
the oscillatory behaviour in $\Delta$ is lost for $\kappa_0>\pi$. 
Similar effect is seen in (b), where oscillations become less prominent beyond this point. Both the results highlight that
system tends to forget the sharp distinction between even $t_\alpha$ and odd $t_\alpha$ because of chaos, which exists from $\kappa_0 \sim 1$ to $3$.

\begin{figure}[h!]
	\centering
\includegraphics[scale=0.35]{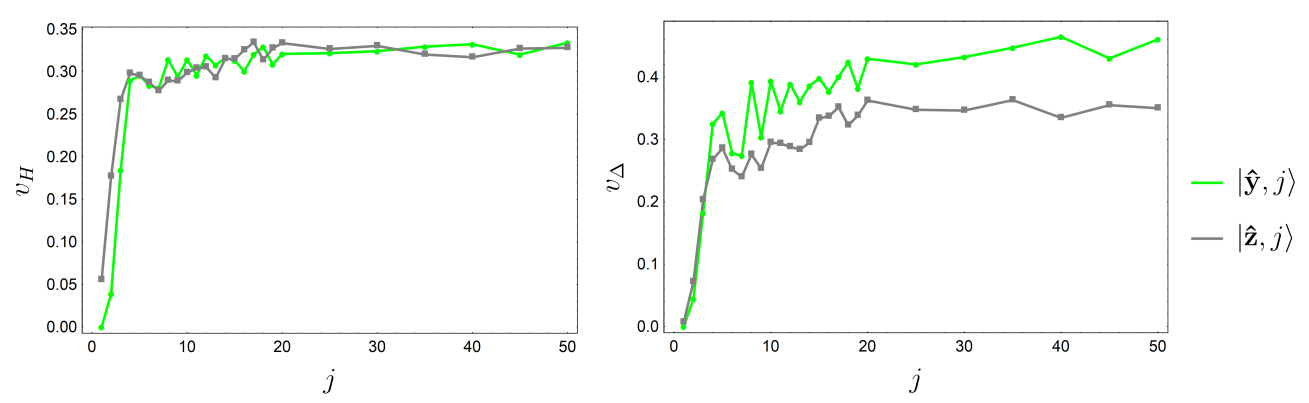}
	\caption{ The time averaged violation  due to measurement for complete chaos) ($\kappa_0=7$) as a function of $j$.
	Here $t_\beta - t_\alpha = n=2$ here, but  this holds for all even $n$ as was 
shown in the case of $j=15$ above.}
	\label{varj}
\end{figure}

The preceding results were all for $j=15$ with 31 states, and it is of interest to see the dependence on this, as the classical limit is effectively reached when $ j \rightarrow \infty$. Shown in Fig.~ \ref{varj} is a scaled quantity $v_\Delta = \expval{\Delta_2(\kappa_0 =7)}/(2j+1) $ and $v_H = \expval{H_2 (\kappa_0=7)}$ as $j$ is varied. The parameter $\kappa_0$ is fixed at a value when classical chaos is dominant, and again the two distinct initial states are examined. Roughly, the degree of violation is seen to increase with $j$ and tends to a constant at surprisingly small values of $j~4-5$. Note that in the kicked top Floquet operator, the torsion term contains the ratio $\kappa_0/j$ and hence for very small $j$, increasing $\kappa_0$ will show periodicity and not reflect classical chaos. Thus for as small angular momenta as $j=3/2$ and $j=2$ the kicked top is exactly solvable 
\cite{ArulTop} but does not reflect the classical dynamics beyond about $\kappa_0=3$, when chaos just sets in. It is well-known that the kicked top can be considered as an all-to-all interacting system of $2j$ qubits, see for example \cite{ghose-pra-2008}. Thus full fledged chaos effects can be felt only for $j$ that is of the order of $\sim 10$,
although it has also been shown that even for small value of $j$ or number of qubits, the system already exhibits some signatures of chaos \cite{PG_2021} or instability such as exponential growth of OTOC, albeit for a very small time.
Thus it seems reasonable that for $\kappa_0=7$, the NSIT measures tend to saturate for $j \approx 10$, although in the $\Delta$ measure it seems to matter what the initial state is. As preliminary results, these indicate that in a closed system with unitary dynamics, the classical limit does not imply that the NSIT will become increasingly valid.

We note that these results are linked consistently with results from quantum chaos.  If one starts from a coherent state, it gets stretched and folded due to the unstable and stable manifolds in a chaotic system, and leads to 
delocalization along with superpositions. This in direct contradiction with the first assumption of macrorealism, because it allows a faster and more prominent superposition between macroscopically distinct quantum states. More quantitatively, quantum chaotic systems have a much earlier time at which quantum effects set in than regular systems: as the Ehrenfest time \cite{Berry_1979,Chirikov_1988,
Logtime_Zurek2002} at which quantum-classical correspondence breaks down scales as $-\log(\hbar)/\lambda$, where $\hbar$ is a scaled Planck constant, scaled by a characteristic action of the system, and $\lambda$ is the Lyapunov exponent, whereas in regular systems this time is much larges and scales $\hbar^{-1/2}$.

	\begin{figure}[h!]
\centering
\begin{subfigure}{.5\textwidth}
  \centering

\includegraphics[width=.9\textwidth]{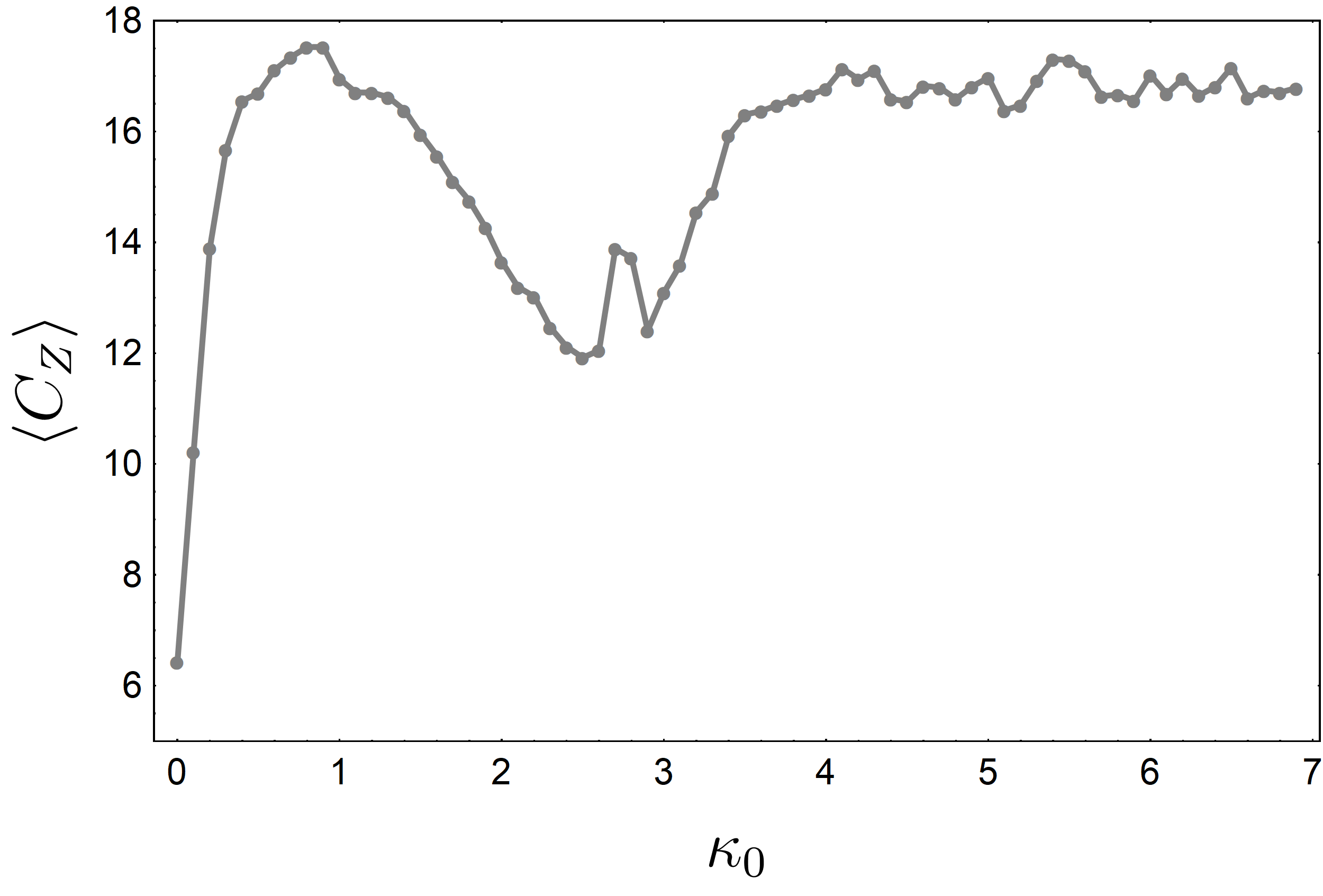}
  \label{fig:sub1}
\end{subfigure}%
\begin{subfigure}{.5\textwidth}
  \centering

\includegraphics[width=.9\textwidth]{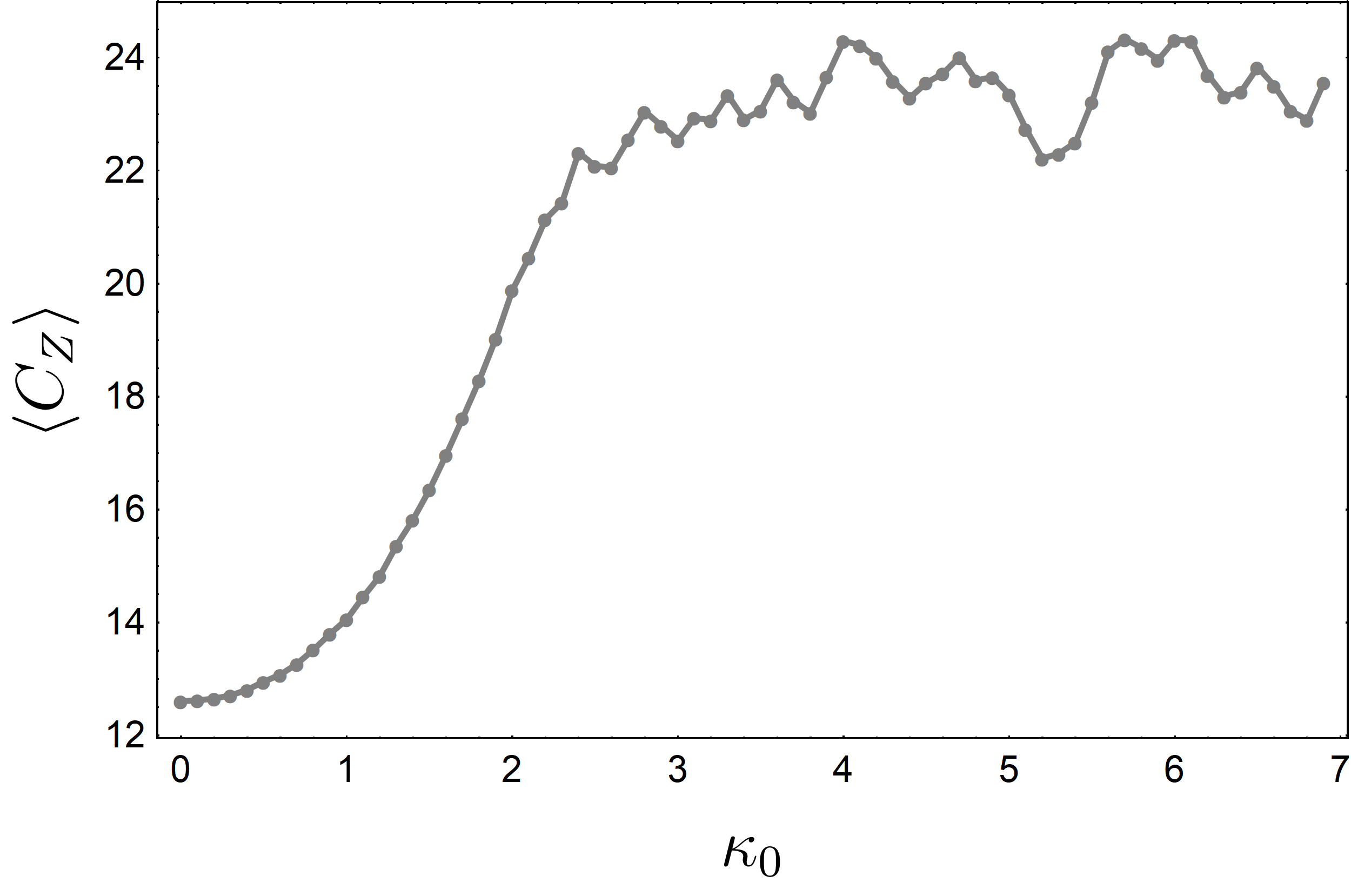}
\end{subfigure}
\caption{The coherence $C_Z\left(\rho  \right) $ in Eq.~(\ref{DistPower}) is shown as an average over the first 50 time steps for $|\mathbf{\hat{z}}, j \rangle$ (left) and  $|\mathbf{\hat{y}}, j \rangle$ (right) initial states. Note the difference in the vertical axis scales between the two figures. The large value of coherence at 
	$\kappa_0=0$ in the second case is due to the large $J_z$ coherence for a $J_y$ eigenstate.}
\label{coherence}
\end{figure}

	\begin{figure}[h!]
\centering
\begin{subfigure}{.5\textwidth}
  \centering
 {\includegraphics[width=.9\textwidth]{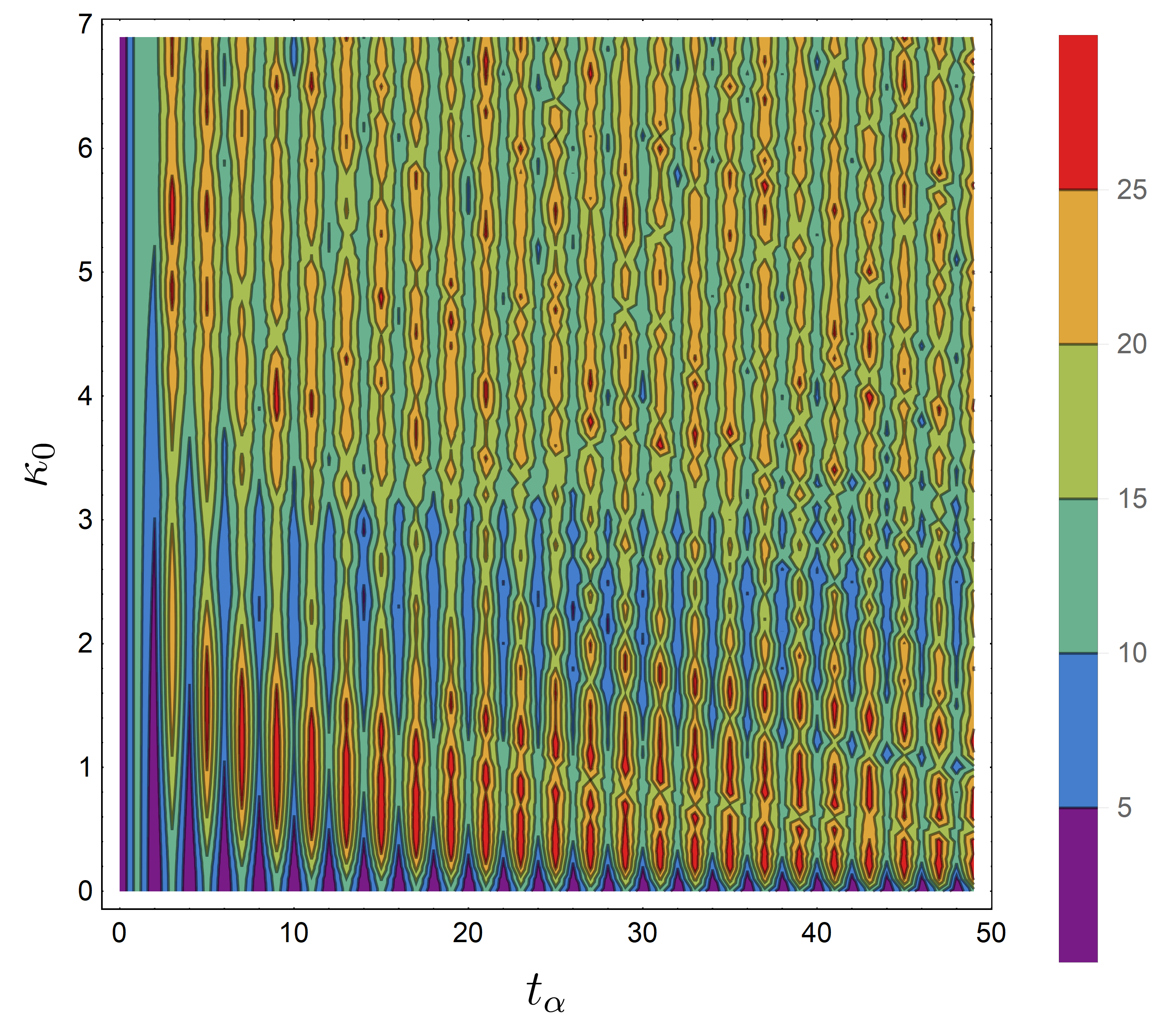}}

  \label{fig:sub1}
\end{subfigure}%
\begin{subfigure}{.5\textwidth}
  \centering {\includegraphics[width=.9\textwidth]{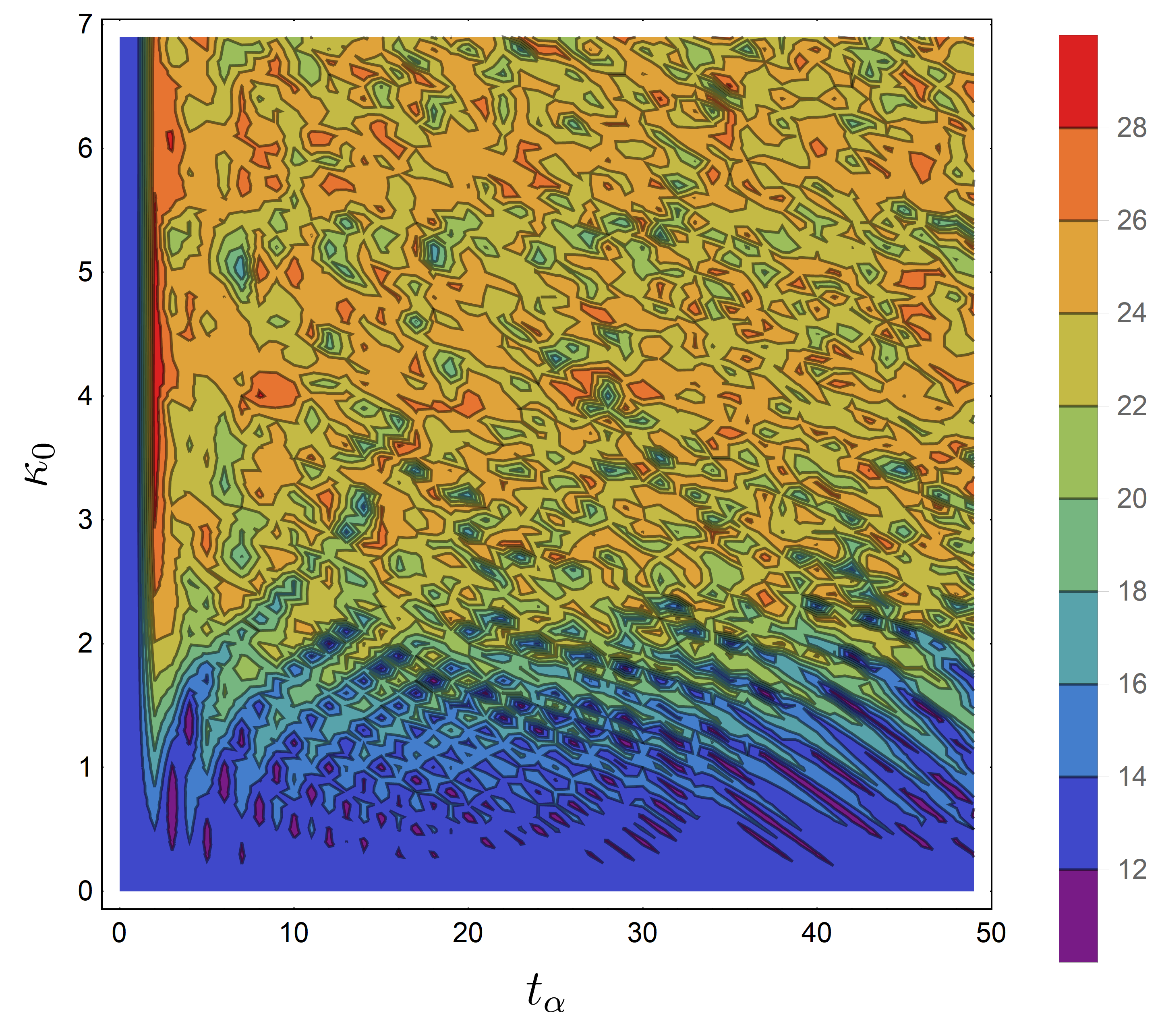}}

  \label{fig:sub2}
\end{subfigure}
\caption{Contour plots of the coherence $C_Z$ as a function of the time at which Alice measures and the parameter $\kappa_0$ are shown here for the two initial states, $|\mathbf{\hat{z}}, j \rangle$ (left) and  $|\mathbf{\hat{y}}, j \rangle$ (right). The correspondence with plots showing NSIT violations as in Fig.~\ref{contour_z} are noted.}
\label{contour_y}
\end{figure}

Finally we calculate the coherence in the state just before Alice's measurement as indicated in Eq.~(\ref{DistPower}. Figure~ \ref{coherence} shows the coherence measure $C_Z$ for the two initial states as a function of the parameter $\kappa_0$. The coherence is found at $0 \leq t_{\alpha} \leq
50$ and averaged.
The 4-period cycle associated with it is stable whenever $(2\cos\kappa_0 + \kappa_0 \sin\kappa_0)^2 <4$. As a result, in the interval $[0, 7]$ there
	are 4 points where the stability of cycle changes. These are $\kappa_0=0, \pi, \sim5.6$ and $2\pi$. Till $\kappa_0 \sim 1$, the cycle
	remains enclosed in a narrow separatrix region, and the at odd times gets predominantly rotated close to a $J_x$ eigenstate which has a large coherence in the $J_z$ basis. But with increasing $\kappa_0$ further, these states have lesser overlaps with $J_z$ eigenstates and leads to a lower coherence. Beyond about $\kappa_0=\pi$, the cycle loses stability and the coherence increases and saturates. Beyond $4$, chaos becomes quite prominent over the whole sphere, and the windows of stability become smaller and does not impact the quantum for these small values of $j$. We notice that in the case of the $J_y$ eigenstate, the coherence shows a smooth and monotonic increase and beyond $\kappa_0=2$ saturates. However, notice that the scales in the two plots are different and hides the large coherences that are present in this second case, simply because of the initial state being uniformly close to that of an $J_y$ eigenstate. This large coherence could be connected with the large violations of the 
    NSIT in these cases, but this needs further investigation.
    The smoothness of the increase also reflects the much simpler classical structure that is far from any separatrix for this initial state. Figure~\ref{contour_y} shows the coherence as a function of time and the parameter $\kappa_0$, without any averaging. These show very similar patterns to those for the NSIT measures in Fig.~\ref{contour_z}, and the influence of coherence on these measures could be explored further, for example as the coherences evolved between the measurements.

\section{Summary and discussions}
We studied violations of macrorealism via the no-signaling in time condition for a paradigm quantum chaotic system, the kicked top. Introducing two measures of 
the violations we have mainly studied numerically their dependence on the chaos parameter and the time between the measurements. We have shown that there is a strong parity effect in the time between measurements that persists quite into the chaotic regimes for certain states. We have shown how the NSIT violations can be very different for different initial states and have also provided qualitative reasoning for these based on semiclassical or classical settings. While there is no simple rule such as larger chaos implies larger NSIT violations, there is a fair amount of correlations and it is true that chaos engenders a large NSIT violation. The role of the stability of periodic orbits in the NSIT violations is clearly seen, and when the classical and quantum dynamics is stable, generic states can lead to smaller values of violation. Although there could be special states that can behave differently. We have seen that these cases have to do with the presence of a large coherence in the measurement basis. Thus these may be exceptional, but again point to a direction for further investigations and connections between quantum coherence and macrorealism conditions.

We have also observed that the NSIT correlations are a form of a
an out-of-time-ordered correlator, a 3-OTOC. The OTOC is a sensitive measure of quantum chaos and instability, and a quantum Lyapunov exponent has been defined with these. Thus the connection is intriguing and deserves more exploration. We highlighted that in a general case, a combination of factors is responsible for the overall violation, one of them being measurement of non-commutative operators in our Alice-Bob setup. Although the details may in general depend on initial state and $n$, the time between measurements, in the presence of chaos we showed that long-time average of disturbances is largely independent of the exact value of $n$ in the chaotic limit. The dynamics can lead to formation of different equivalent sets of $n$ values, within which these differences almost completely vanish. Finally,  the variation of disturbance as a function of $j$ shows that effects of chaos are more potent for larger $j$, which implies that chaotic systems are the among the best places to study properties such as macroscopic coherence and tests of macrorealism. 

\begin{acknowledgments}
We would like to thank Dipankar Home for generous email discussions on an earlier version of this work.
\end{acknowledgments}

\section*{Appendix: Some aspects of classical and quantum dynamics of the kicked top}

 Here we review the classical limit, focusing on the chosen initial states, followed by study of their quantum evolution. The latter explains how an 
 increase in $\kappa_0$ leads to an increase in $C_{A}(\rho)$ for localized states.

\subsection{Classical Map}
Taking the limit of 
$j\ra \infty$ in \eqref{Hamiltonian} with $J = \sqrt{j(j+1)}$ gives us the
classical Hamiltonian. If we think of the classical vector $\mathbf{J}$ in spherical polar coordinates, so that\begin{equation}\label{key}
\mathbf{J}/J = (\sin\theta\cos\phi, \sin\theta\sin\phi,\cos\theta),
\end{equation} solving the Hamilton's equations (taking $q = \phi$, $p = \cos\theta$) it is easily seen that the $\delta$ kick is an impulsive rotation 
of $\vb{J}$ about $z$ axis by an angle $J_z\kappa_0$,
where $J_z = J \cos\theta$. During the kick, the rotation bit of Hamiltonian is negligible and thus Hamilton's equation integrates to give \begin{equation}\label{phirot}
\phi(n^+)- \phi(n^-) = \int_{n^-}^{n^+} dt \dot\phi = \int_{n^-}^{n^+} dt \pdv{H}{p} = J \kappa_0 \cos\theta \int_{n^-}^{n^+} dt \sum_{-\infty}^{\infty} \delta (t - k)  = J_z \kappa_0 
\end{equation}
for any $n$. Therefore, we see that this system is rotating about 
two axes in each turn; by a constant angle of $\frac{\pi}{2}$ around the 
$y$ axis, and by a variable angle around the z-axis. This ``variation" 
in the angle is a necessary ingredient of chaos. 

This system evolves according to a 2D map -  because $\mathbf{J}^2 $ is a 
constant of the motion - which is
\begin{align}\label{key}
	X_i &= Z_{i-1}\cos{(\kappa_0X_{i-1})} + Y_{i-1}\sin(\kappa_0 X_{i-1})\\
	Y_i &= -Z_{i-1}\sin(\kappa_0 X_{i-1})+Y_{i-1}\cos{(\kappa_0 X_{i-1})}\\
	Z_i &= -X_{i-1}
\end{align}

where $X,Y,Z = J_{x,y,z}/j$, and obey $X^2+Y^2+Z^2 = 1$ \cite{Haake1987}. These equations can be obtained from Heisenberg's equations of motion in the classical limit. 
The fixed points at the poles $Y=\pm 1$ and the equitorial 4 period 
cycle $Z = 1 \to X = 1 \to  Z = -1 \to X = -1 $ are of special relevance to us; 
each of these exists for all $\kappa_0$ values. In the quantum case, these fixed points and this cycle correspond to $|\mathbf{\hat{y} }, \pm j  \rangle $ and
$| \mathbf{\hat{z}}, j \rangle$ respectively.

\begin{figure}
	\centering
	\begin{subfigure}{.5\textwidth}
		\centering
		{\includegraphics[width=.9\textwidth]{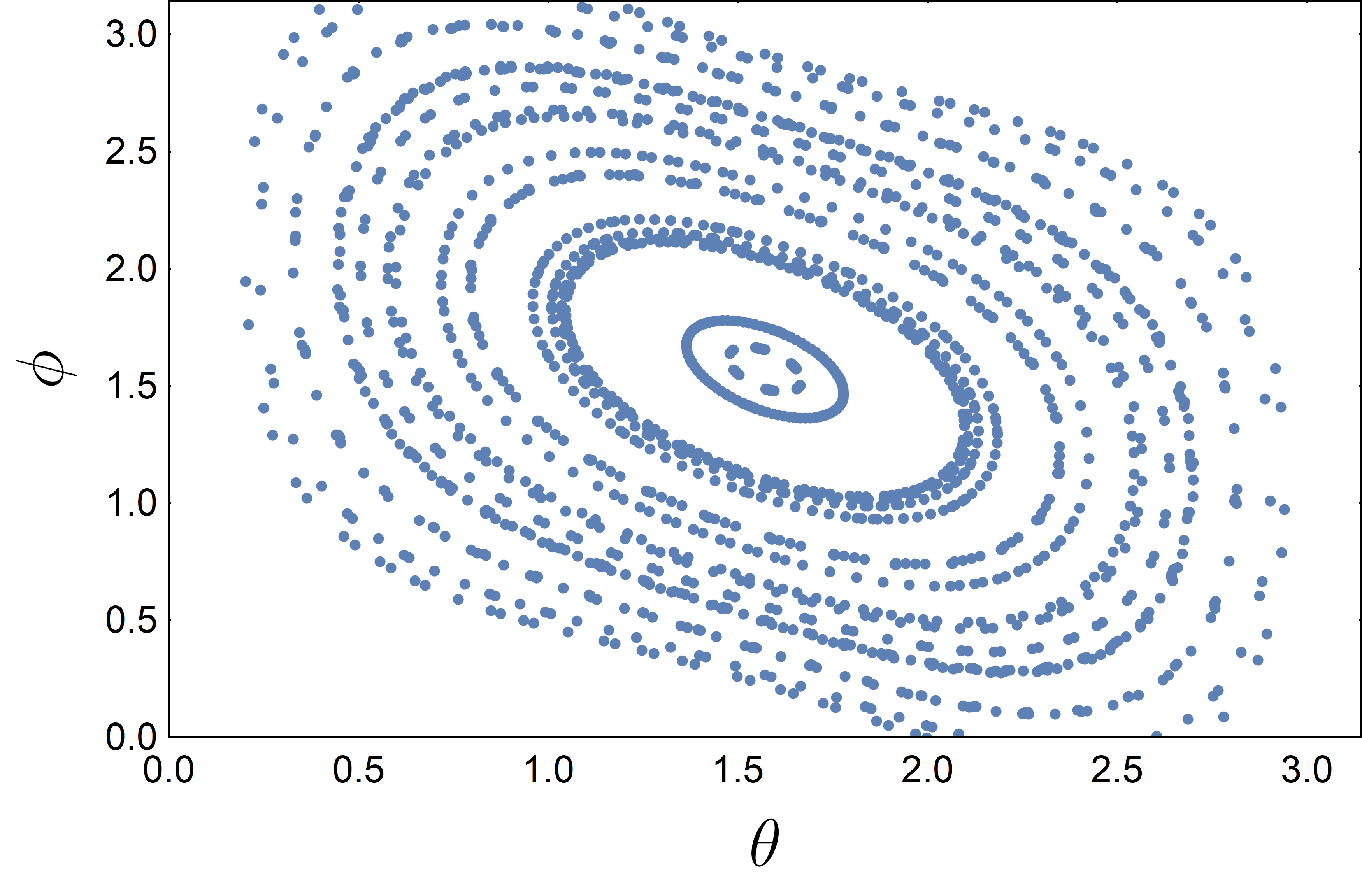}}
		
		\caption{$\kappa_0=1$}
		\label{fig:sub1}
	\end{subfigure}%
	\begin{subfigure}{.5\textwidth}
		\centering {\includegraphics[width=.9\textwidth]{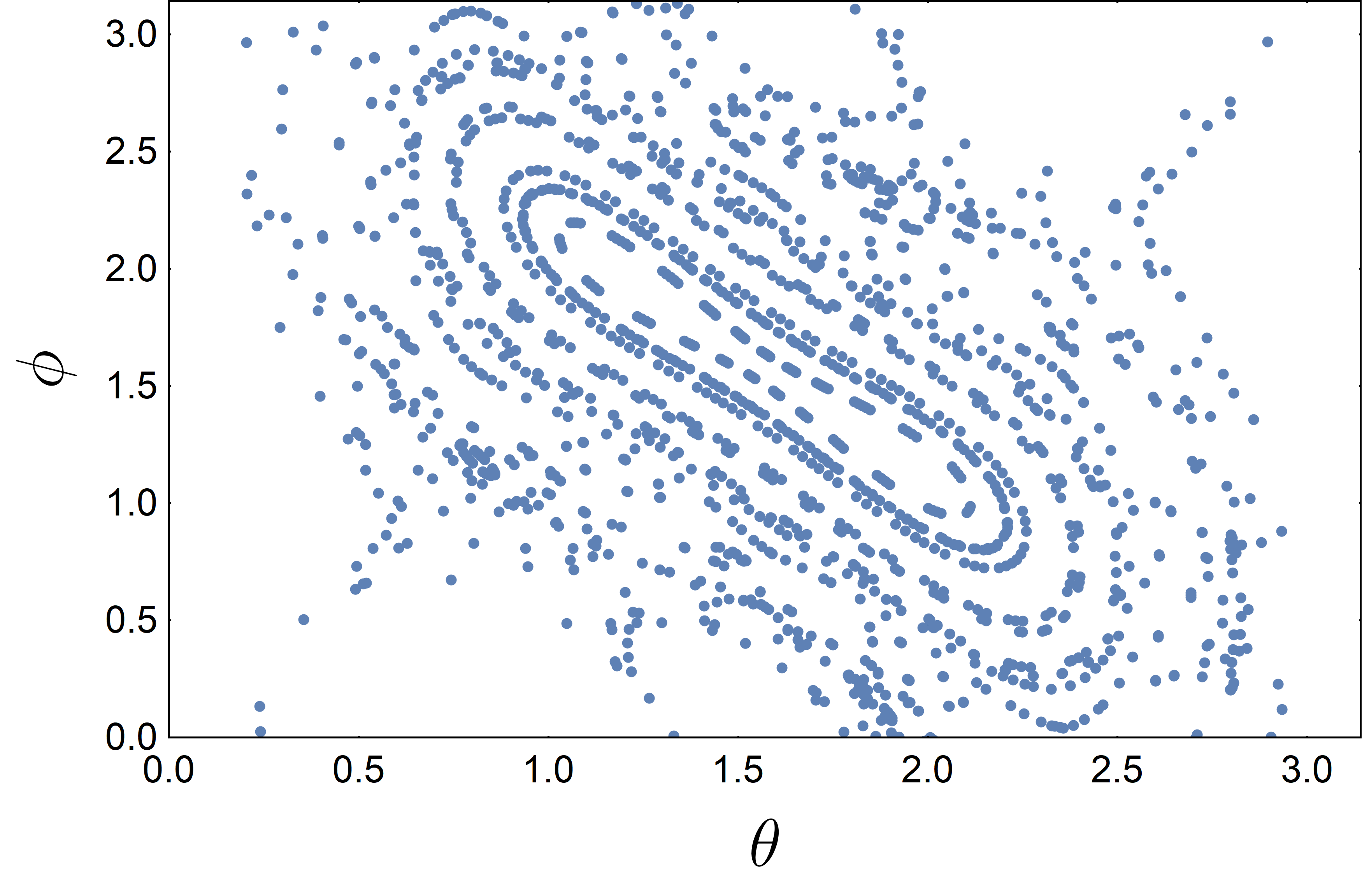}}
		
		\caption{$\kappa_0=2.2$}
		\label{fig:sub2}
	\end{subfigure}
	\caption{Phase space trajectories in the vicinity of fixed point at $Y=1$ for two different cases, on the left, motion is quite regular whereas on the right, the bifurcation at $\kappa_0=2$ has created instability. }
	\label{phasespace}
\end{figure}

As $\kappa_0$ increases, new fixed points and cycles are born, and on further increase, they become 
unstable, to bifurcate into new fixed points and cycles. The game starts at $\kappa_0 = 2$, when the fixed points at $Y=\pm 1$ lose their stability, giving 
rise to two new FPs. Figure \ref{phasespace} shows the trajectories near $Y=1$ as $\kappa_0$ changes from 1 to 2.2.  By $\kappa_0=3$, most of the sphere becomes chaotic, but there are some significant islands of stability, notably the ones around new 
fixed points and one around the 4 period cycle mentioned above, which loses stability at $\pi$; it is stable whenever $(2\cos\kappa_0 + \kappa_0 \sin\kappa_0)^2 <4$. At $\sqrt{2} \pi\sim 0.442$, these FPs become unstable as well.  
By $\kappa_0 = 6$, system is essentially fully chaotic. See \cite{Haake1987, Haake} for 
details. 


\subsection{Quantum Dynamics}


\subsubsection{Dynamics for initial state $ \ket{\vb{\hat z}, j} $}
We will find it convenient to move back and forth between the $J_z$ and $J_x$ basis. From \eqref{Map} it follows that the projectors 
$Z_m = \ket{\vb{\hat{z}},m}\bra{\vb{\hat{z}},m}$ and $X_m =\ket{\vb{\hat{x}},m}\bra{\vb{\hat{x}},m}$   obey

\begin{equation}\label{key}
T Z_m T^{-1} = Z_m; \quad R^2 Z_m R^{-2} = Z_{-m}; \quad R Z_m R^{-1} = X_m; \quad R X_m R^{-1} = Z_{-m}.
\end{equation}
Note that for $\kappa_0 = 0$, the dynamics is trivial. We have a rotating vector starting on $Z_j$, which goes to the $J_x$ eigenstate $X_j$, 
followed by $Z_{-j}$ and finally back to $X_j$, completing the cycle. 
For small but non-zero $\kappa_0$, $T\neq 1$. Simplifying $T X_j T^{-1}$ by using Baker-Hausdorff formula, \cite{sakurai}
\begin{equation}\label{Hadamard}
T X_j T^{-1} = X_j - \frac{i \kappa_0 }{2j}[J_z^2, X_j] - \frac{\kappa_0^2}{4j^2}[J_z^2, [J_z^2,X_j ]]  \cdots.
\end{equation}
Using $K_{\pm} = J_y \pm i J_z$, in analogy to standard raising and lowering operators, after some calculations, one finds \begin{equation}\label{star}
[J_z^2, X_j] =  \frac{1}{2}\sqrt{j\left( 2j-1 \right) }  ( \op{\vb{\hat{x}},j}{\vb{\hat{x}},j-2} - \op{\vb{\hat{x}},j-2}{\vb{\hat{x}},j})
\end{equation}

and $[J_z^2, [J_z^2, X_j]]$ carries terms like $X_{j-2}, X_{j}, |\mathbf{\hat{x}},j-2 \rangle   \langle  \mathbf{\hat{x}}, j|, |\mathbf{x}, j \rangle \langle  \mathbf{\hat{x}}, j-4|$.

For small $\kappa_0$, we may neglect the higher order terms, to conclude that operation of $T$ produces some off diagonal terms and 
mixes $X_j$ with nearby states. 
Because the ``ladder" is at an end for $m=j$, we only got the lower state $X_{j-2}$, but for $m \neq \pm j$, we do get neighbours on both sides. Note that the 
immediate neighbours are not mixed in the process.

After a rotation, $X_m \rightarrow Z_{-m}$. The off-diagonal $J_x$ terms also
rotate, in a sense. Note that $R^4 =\pm 1$, 
and the fact that $R$ cannot distinguish between $z$ axis and $x$ axis. It can only do a counter-clock rotation by $\pi/2$ of each of their eigenstates, 
giving the same phase $\phi_{m}$ for $X_{\pm}$ and $Z_{\pm}$, if any.
Therefore,
\begin{equation}
R|\mathbf{\hat{x}},m \rangle \langle  \mathbf{\hat{x}},n| R^{-1} = 
\phi_{mn}|\mathbf{\hat{z}},m \rangle 
\langle  \mathbf{\hat{z}},n|. \end{equation}
Torsion does nothing to the $Z$'s, whereas the 
off-diagonal $J_z$ terms again pick up opposite phases, but clearly, the 
magnitude of the off-diagonal is not affected. This explains how
the coherence \eqref{DistPower}  increases in both $J_z$ and $J_x$ basis with increase in $\kappa_0$.

As $\kappa_0$ increases, the higher order terms become relevant and as a result the mixing becomes stronger. For such cases, a single kick can mix several 
$\{X_m\}$ states.

\subsubsection{ Dynamics for $\ket{ \vb{\hat y}, j} $}	
Defining $Y_m = \op{\vb{\hat y}, m}$, it follows from \eqref{Map} that
\begin{equation}\label{key}
R Y_m R^{-1} = Y_{m}; \quad \bar{R} X_m \bar{R}^{-1} = Y_m; \quad T Y_m T^{-1} = \bar{R}(TX_m T^{-1})\bar{R}^{-1}
\end{equation} 
where $\bar{R} = \exp(-iJ_z \pi/2)$. For $\kappa_0 = 0$, the state is invariant because it is an eigenstate of $R$. For $\kappa_0>0$, it is clear that behavior is similar to what we had before. 
The action of $T$ on $Y_m$ is just like its action on $X_m$'s, and
mixes neighbouring states in $J_y$ basis too. Of course, this 
should be expected from symmetry  between  $x$ and  $y$ axes with respect to  $z$ axis.

	\bibliographystyle{unsrt}
	\bibliography{quantumchaos.bib}

\end{document}